%% file: Dx_eprint.tex
\documentclass[prl,showpacs,superscriptaddress,floatfix,12pt]{revtex4}

\usepackage{graphics}  
\usepackage{makeidx}
\usepackage{colordvi}
\begin{document}  

\title{Measurement of Prompt Charm Meson Production Cross Sections
in $p\bar{p}$ Collisions at $\sqrt{s}=1.96\,\mbox{TeV}$}
 
\input Limited_Run2_Auth_revtex4.tex

\begin{abstract} 
We report on measurements of differential cross sections $d\sigma/dp_T$ for prompt
charm meson production in $p \bar{p}$ collisions at $\sqrt s=1.96$\,TeV
using $5.8\pm 0.3\,\mbox{pb}^{-1}$ of data from the CDF\,II detector at the Fermilab 
Tevatron.
The data are collected with a new trigger that  is sensitive to 
the long lifetime of hadrons containing heavy flavor.
The charm meson cross sections are measured 
in the central rapidity region $|y|\leq1$
in four fully reconstructed decay modes:
$D^0\to K^-\pi^+$, $D^{*+}\to D^0\pi^+$, 
$D^+\to K^-\pi^+\pi^+$, $D_s^+\to \phi\pi^+$, and their charge conjugates.
The measured cross sections are compared to theoretical calculations.
\end{abstract}
 
\maketitle    

Measurements of the production cross~sections of hadrons containing
$b$~quarks or charm~quarks (heavy flavor hadrons)
in $p \bar{p}$ collisions provide an opportunity
to test predictions based on Quantum Chromodynamics (QCD).
Previous measurements of $B$~meson production cross sections in
$p\bar{p}$ collisions at $\sqrt s=1.8$\,TeV~\cite{Abbott:1999se,Acosta:2001rz} 
were about three times 
larger than Next-to-Leading-Order (NLO) QCD predictions~\cite{QCDcalc},
although recent calculations with a more accurate description of
$b$~quark fragmentation have reduced this discrepancy to a
factor~1.7~\cite{Binnewies:1998vm,Cacciari:2002pa}.
Charm meson production cross~sections have not been measured
in $p\bar{p}$ collisions and may help with understanding this disagreement.
The upgraded Collider Detector at Fermilab (CDF\,II)
has a new capability to trigger on tracks displaced from the
beamline originating from the
decay of long-lived hadrons containing heavy flavor quarks.
We report here measurements of prompt charm meson cross sections
using data recorded with this trigger
in February and March 2002, corresponding to $5.8\pm0.3\,\mbox{pb}^{-1}$ of integrated luminosity.

An overview of the CDF\,II detector can be found elsewhere~\cite{Blair:1996kx},
only the components relevant to this analysis are described here.
The CDF coordinate system has the $z$~axis pointing along the proton momentum;
$\varphi$ is the azimuth, $\theta$ is the polar angle,
and $r$ is the distance from the proton beam axis.
The CDF\,II central tracking region covers the pseudorapidity region $|\eta|\leq1$,
where $\eta=-\ln[\tan(\theta/2)]$.
A superconducting solenoid provides a nearly uniform axial field of 1.4\,T.
The silicon vertex detector~(SVX\,II)~\cite{Sill:zz} consists of double-sided 
microstrip sensors arranged in five concentric cylindrical shells with radii 
between 2.5 and 10.6\,cm.
Surrounding the SVX\,II is the Central Outer Tracker (COT)~\cite{Pitts:qy},
an open cell drift chamber covering radii from 40 to 137 cm.
The COT has 96 layers,
organized in 8 superlayers, alternating between axial and $\pm2^\circ$ stereo readout.

CDF\,II has a three-level trigger system. 
We describe here the trigger used in this analysis.
At the first trigger level, charged tracks are reconstructed in the COT axial projection
by a hardware processor (XFT)~\cite{Thomson:2002xp}.
The trigger for hadronic charm decays requires two oppositely
charged tracks with $p_T\ge 2\,\mbox{GeV}/c$ and
the scalar sum of the $p_T$'s larger than 5.5\,GeV/$c$,
where $p_T$ is the magnitude of the component of the momentum transverse to the beam axis.
At the second trigger level, the Silicon Vertex Tracker (SVT)~\cite{Ashmanskas:1999ze}
associates axial strip clusters from the four inner SVX\,II layers with XFT track information.
The SVT measures the distance of closest approach of a track relative to the
beam axis (impact parameter or $d_0$) with a resolution of 50\,$\mu$m,
which includes a contribution of 30\,$\mu$m from the beam spot transverse size.
Events containing hadronic decays of heavy flavor hadrons 
are selected by requiring two tracks with 
$120\,\mu\mbox{m}\leq d_0 \leq1\,\mbox{mm}$ each.
At the third trigger level, a farm of computers performs complete event reconstruction online;
the opening angle $\delta\varphi$ between the two trigger tracks is required to be
between 2$^\circ$ and 90$^\circ$,
and the intersection point in the $r$-$\varphi$ plane
projected along the net momentum vector of the two tracks
must be more than 200\,$\mu$m from the beamline.

\begin{figure}[p]
\scalebox{1}{\includegraphics{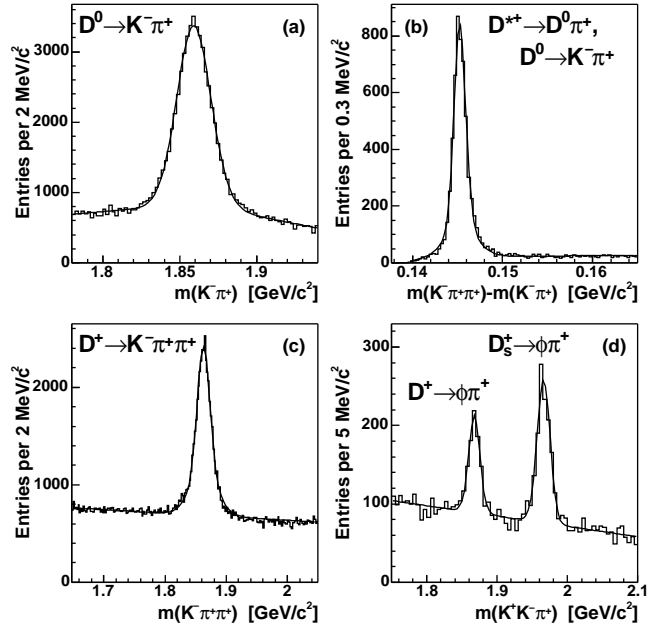}}
\caption{\label{fig:signal} 
Charm signals summed over all $p_T$ bins: 
(a) invariant mass distribution of  $D^0\to K^-\pi^+$ candidates; 
(b) mass difference distribution of $D^{*+}\to D^0 \pi^+$ candidates;
(c) invariant mass distribution of $D^+\to K^-\pi^+\pi^+$ candidates;
(d) invariant mass distribution of $D^+\to\phi\pi^+$ and $D_s^+\to\phi\pi^+$
candidates.
The curves show the results of the fits described in the text.
}
\end{figure}

We reconstruct charm mesons in the following decay modes:
\mbox{$D^0\to K^-\pi^+$}, 
\mbox{$D^{*+}\to D^0\pi^+$} with \mbox{$D^0\to K^-\pi^+$},
\mbox{$D^+\to K^-\pi^+\pi^+$}, 
\mbox{$D_s^+\to \phi\pi^+$} with \mbox{$\phi \to K^+K^-$},
and their charge conjugates.
For every track pair that satisfies the trigger requirements (trigger pair),
we form one $D^0\to K^-\pi^+$ candidate
and a second candidate with the mass assignments swapped.
No particle identification is used in this analysis.
$D^0$ candidates within three standard deviations of the $D^0$ mass 
are combined with a third track with $p_T\geq0.5\,$GeV/$c$ to form 
$D^{*+}\to D^0 \pi^+$ candidates.
The three-body decays of the $D^+$ and $D^+_s$ are reconstructed
by combining a trigger pair with a third track having axial hits 
in at least three out of five SVX\,II layers
and performing a vertex fit based on axial track information only.
For $D^+_s$ reconstruction, we specifically require
the $K^-\pi^+$ pair to satisfy the trigger requirements,
since the typical opening angle between two kaons from $\phi$ decay
is close to the $\delta\varphi\geq2^\circ$ trigger requirement.
Each $\phi$ candidate is required to have a mass within $\pm$20\,MeV of
the world average $\phi$ mass~\cite{Hagiwara:fs}.
The $D$ meson candidates are binned in $p_T$ 
as indicated in Table~\ref{table_Dx_D_bincorr}.
The signals summed over all $p_T$ bins are shown in Fig.~\ref{fig:signal}.

The $D^0$ yield is obtained from a binned maximum likelihood fit
to the $K^-\pi^+$ invariant mass distribution,
with a linear function for the combinatoric background,
a narrow Gaussian for the $D^0$ signal,
and a wide Gaussian with the same normalization describing $D^0\to K^-\pi^+$ with the wrong mass assignment.
We determine the shape of the mass distribution resulting from the wrong mass 
assignment using $D^0$
from $D^{*+}$ decay where the charge of the low-momentum pion
determines the mass assignment of the $D^0$ decay products. 
The $D^{*+}$ yield is extracted from the distribution of 
$\Delta m=m(K^-\pi^+\pi^+)-m(K^-\pi^+)$, 
the mass difference between 
the $K^-\pi^+\pi^+$ and the $K^-\pi^+$ combination. 
The signal is modeled with two Gaussians with equal means,
and the background is characterized as 
$a \sqrt{\Delta m-m_\pi}\exp(b(\Delta m-m_\pi))$,
where $a$ and $b$ are free parameters in the fit.
The $D^+$ signal is described with two Gaussians
and the background is described with a linear function.
In the $\phi \pi^+$ mode, we model the invariant mass distribution
as a linear background with two Gaussians, one corresponding to the $D^+$
and one to the $D_s^+$.
We find $36804\pm409$ $D^0\to K^-\pi^+$, $5515\pm85$ $D^{*+}\to D^0 \pi^+$,
$28361\pm294$ $D^+\to K^-\pi^+\pi^+$, and $851\pm43$ $D_s^+ \to \phi \pi^+$,
where the uncertainties quoted are statistical only.
We vary the signal and background models and attribute 
systematic uncertainties on the signal yield in the range of $1\%-6\%$,
depending on the decay mode and the $p_T$ range of the candidates.

\begin{figure}[p]
\scalebox{1}{\includegraphics{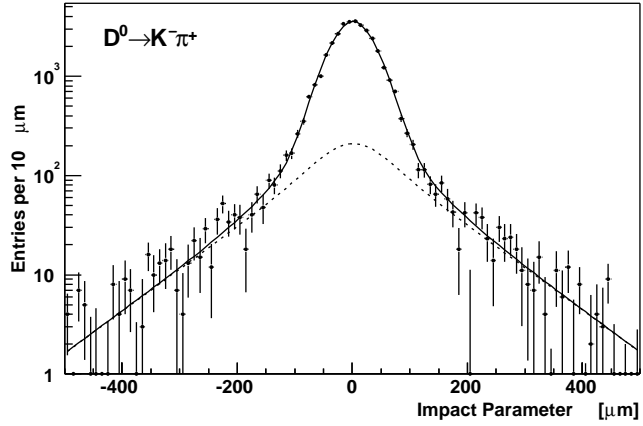}}
\caption{\label{fig:d0} 
The impact parameter distributions of the $D^0$ mesons,
measured from the $\pm 2\sigma$ signal region of the invariant mass distribution and
corrected for combinatoric background measured in the invariant mass sidebands.
The solid curve is the fit result summed over all $p_T$ bins.
The dashed curve shows the contribution of secondary charm from $B$ decay.
}
\end{figure}

We separate charm directly produced in $p\bar{p}$ interactions (prompt charm)
from charm from $B$ decay (secondary charm) 
using the impact parameter of the net momentum vector
of the charm candidate to the beamline~\cite{Chen:2003qe}.
Prompt charm mesons point to the beamline.
The shape of the impact parameter distribution of secondary charm is obtained from a 
generator-level NLO Monte Carlo (MC) simulation
of $B$ meson production~\cite{Acosta:2001rz} and decay~\cite{qq},
smeared with a resolution function (Gaussian + exponential tails) 
obtained from a sample of $K^0_S\to\pi^+\pi^-$ decays
that satisfy the trigger requirements.
The impact parameter distribution of the reconstructed charm samples,
shown for the $D^0$ in Fig.~\ref{fig:d0}, is fit to a prompt and a secondary component. 
The prompt fraction is measured for each $p_T$ bin.
Averaged over all $p_T$ bins, 
$(86.6\pm0.4)\%$ of the $D^0$ mesons,
$(88.1\pm1.1)\%$ of $D^{*+}$,
$(89.1\pm0.4)\%$ of $D^+$, and
$(77.3\pm3.8)\%$ of $D_s^+$ are promptly produced (statistical uncertainties only).
The systematic uncertainties on the prompt fractions are estimated by 
removing the non-Gaussian tail in the resolution function 
and evaluating the variation.
The relative uncertainty is found to be in the $3\%-4\%$ range, depending on the decay mode.

Using a hit-level simulation of the COT, overlaid with data events from the
hadronic heavy flavor trigger to reproduce a realistic occupancy,
we measure a reconstruction efficiency in the COT of $99\%$ for tracks with 
$p_T\ge 2\,\mbox{GeV}/c$, falling to $95\%$ at $p_T=0.5\,\mbox{GeV}/c$.
The efficiency for finding three SVX\,II axial hits on a reconstructed track is
measured from data to be about $85\%$.
The efficiencies of the trigger hardware XFT and SVT to reconstruct tracks
are measured from data samples without XFT or SVT requirements~\cite{Chen:2003qe}.
The XFT tracking efficiency is greater than $95\%$.
In the data-taking period considered,
the SVX\,II and the SVT were not yet fully operational,
and the efficiency varied as certain SVX\,II modules were
included or excluded from the trigger.
Therefore, we measured the SVT efficiency in 42 periods, each corresponding to one 
$p\bar{p}$~store of the Tevatron, and characterized the efficiency
as a function of the track azimuth~$\varphi$, the longitudinal
position $z_0$, the polar angle $\theta$ and the transverse momentum $p_T$.
The average single-track efficiency of the SVT for this early data taking period was about 42\%.

The measured efficiencies are applied to 
a generator-level NLO MC simulation of charm meson production and decay
to calculate the trigger and reconstruction efficiencies,
taking into account decay in flight and hadronic interactions
of the charm meson decay products.
The MC $p_T$ spectrum of the charm mesons is reweighted to match the measured $p_T$ spectrum.
The integrated cross section $\sigma_i$ in each $p_T$ bin $i$ with $|y|\le 1$ 
(where $y=\frac{1}{2}\ln(\frac{E+p_z}{E-p_z})$ and $E$ is the energy of the charm meson)
is calculated using the following equation:
\begin{equation}
\sigma_i=\frac{N_i/2\cdot f_{D,i}}{\int\mathcal{L}dt\cdot\epsilon_i\cdot {\cal B}},
\end{equation}
where $N_i$ is the number of charm mesons in each $p_T$ bin and 
$f_{D,i}$ is the fraction of prompt charm in that bin.
The integrated luminosity $\int\mathcal{L}dt$ at CDF is 
normalized to an inelastic cross section of
$\sigma_{p\bar{p}}=60.7\pm2.4$\,mb~\cite{inelastic}.
The rate of inelastic colisions is measured with
Cherenkov luminosity counters~\cite{Lerr} 
and has an uncertainty of 4.4\%.
The factor $\frac{1}{2}$ is included because we 
count both $D$ and $\bar{D}$ mesons,
but we report cross sections for $D$ alone.
We verified that the $D$ and $\bar{D}$ cross sections are equal 
within statistical uncertainties.
The branching fractions ${\cal B}$ are taken from Ref.~\cite{Hagiwara:fs}.
For the $D^0$ cross section, 
we sum the branching fractions of $D^0\to K^-\pi^+$ and $D^0\to K^+\pi^-$,
since both contribute to the observed signal.
The combined reconstruction and trigger efficiency $\epsilon_i$ varies from $0.12\%$
to $1.9\%$ depending on the decay mode and the $p_T$ bin.
Systematic uncertainties
on the trigger and reconstruction efficiency arise predominantly from
the uncertainty on single-track efficiencies
and two-track efficiency correlations. 
They also have contributions from ionization energy loss,
hadronic interactions in the inner tracker material
and the size of the interaction region.
The combined systematic uncertainty on the trigger and reconstruction efficiency
is in the range of $8\%-14\%$, depending on the decay mode and the $p_T$ 
range of the $D$ mesons.

The total cross sections are obtained by summing over all $p_T$ bins.
However, the last $p_T$ bin is replaced by an inclusive bin with $p_T > 12$\,GeV/$c$.
We find $\sigma(D^0, p_T\geq 5.5\,\mbox{GeV}/c, |y|\le1)=13.3\pm0.2\pm 1.5\,\mu$b,
$\sigma(D^{*+}, p_T\geq 6.0\,\mbox{GeV}/c, |y|\le1)=5.2\pm0.1\pm 0.8\,\mu$b,
$\sigma(D^+, p_T\geq 6.0\,\mbox{GeV}/c, |y|\le1)=4.3\pm0.1\pm 0.7\,\mu$b and
$\sigma(D_s^+, p_T\geq 8.0\,\mbox{GeV}/c, |y|\le1)=0.75\pm0.05\pm 0.22\,\mu$b,
where the first uncertainty is statistical and the second systematic.
To calculate the differential cross sections, we divide $\sigma_i$
by the width of the $p_T$ bin.
Since we report $d\sigma/dp_T$ at the center of each $p_T$ bin,
we apply a correction to account for the non-linear shape of the cross section,
using the $p_T$ reweighted MC to obtain the shape of the cross-section inside each $p_T$ bin.
The results are
listed in Table~\ref{table_Dx_D_bincorr}.

\begin{figure}[p]
\scalebox{1}{\includegraphics{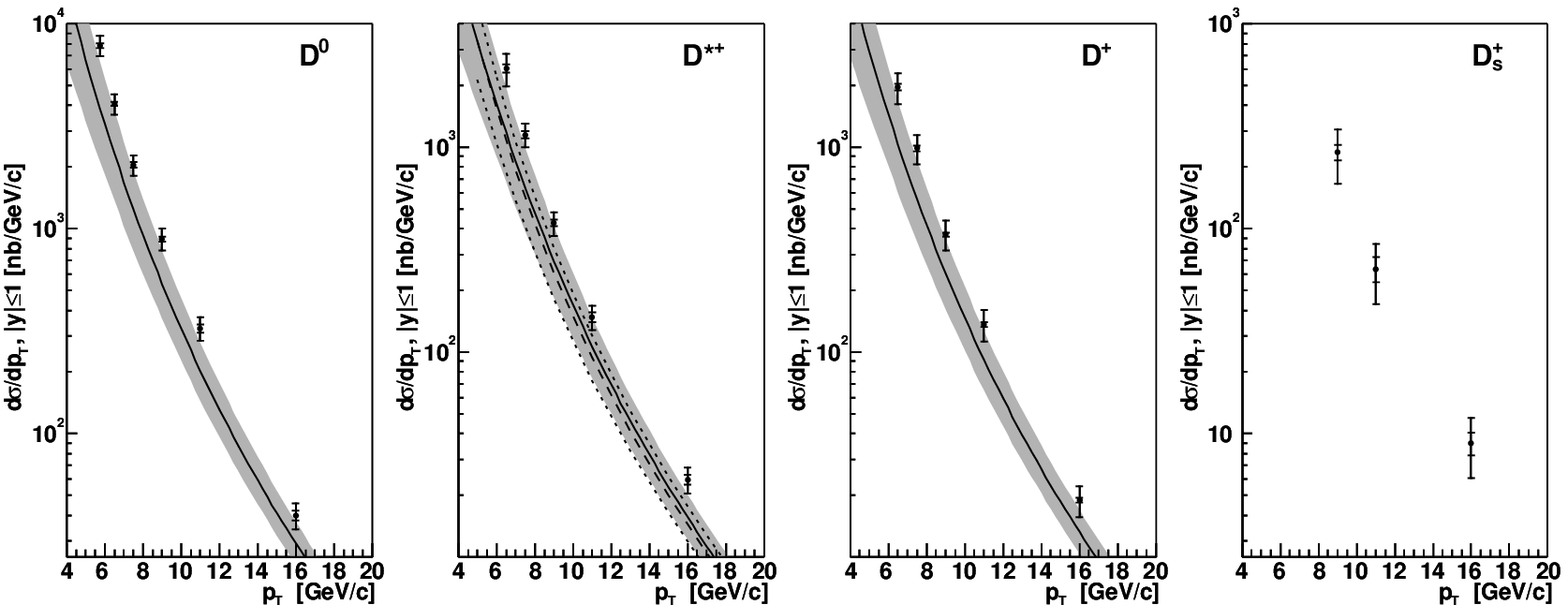}}
\caption{\label{fig:Dx} 
The measured differential cross section measurements for $|y|\leq1$, shown by the points.
The inner bars represent the statistical uncertainties; the outer bars are the 
quadratic sums of the statistical and systematic uncertainties.
The solid curves are the theoretical predictions from Cacciari and Nason~\cite{Cacciari:2003zu},
with the uncertainties indicated by the shaded bands.
The dashed curve shown with the $D^{*+}$ cross section is the theoretical prediction 
from Kniehl~\cite{kniehl};
the dotted lines indicate the uncertainty.
No prediction is available yet for $D_s^+$ production.
}
\end{figure}

The measured differential cross sections are compared to two recent calculations~\cite{Cacciari:2003zu,kniehl},
as shown in Fig.~\ref{fig:Dx}.
The uncertainties on the calculated cross sections are evaluated by varying 
independently the renormalization and factorization scales
between 0.5 and 2 times the default scale.
Ref.~\cite{Cacciari:2003zu} uses a default scale of $\sqrt{m^2_c+p^2_T}$,
where $m_c = 1.5$\,GeV/$c^2$ is the $c$ quark mass,
while Ref.~\cite{kniehl} uses a default scale of $2\sqrt{m^2_c+p^2_T}$.
Contributions from other sources, such as the charm quark mass, the value of the strong coupling constant
and the fragmentation functions, were reported to be smaller and are not taken into account.

In conclusion, the measured differential cross sections are higher
than the theoretical predictions
by about 100\% at low $p_T$ and 50\% at high $p_T$. 
However, they are compatible within uncertainties.
The same models also underestimate $B$ meson production
at $\sqrt s=1.8\,$TeV by similar factors~\cite{Acosta:2001rz,Binnewies:1998vm,Cacciari:2002pa}.

\begin{acknowledgments}   
We thank the Fermilab staff and the technical staffs of the
participating institutions for their vital contributions.  This work was
supported by the U.S. Department of Energy and National Science
Foundation; the Italian Istituto Nazionale di Fisica Nucleare; the
Ministry of Education, Culture, Sports, Science and Technology of Japan;
the Natural Sciences and Engineering Research Council of Canada; the
National Science Council of the Republic of China; the Swiss National
Science Foundation; the A.P. Sloan Foundation; the Bundesministerium fuer
Bildung und Forschung, Germany; the Korean Science and Engineering
Foundation and the Korean Research Foundation; the Particle Physics and
Astronomy Research Council and the Royal Society, UK; the Russian
Foundation for Basic Research; and the Comision Interministerial de
Ciencia y Tecnologia, Spain.
\end{acknowledgments}

\newpage
\begin{table}[p]
\begin{ruledtabular}      
\begin{tabular}{cccccc}
 & & \multicolumn{4}{c}{$d\sigma(|y|\le1)/d p_T\;\;\;\mbox{[nb/(GeV/c)]}$}\\\hline   
 $p_T\;\mbox{range}$ & $\mbox{Central}\;p_T$ & $D^0$ & $D^{*+}$ & $D^+$ & $D^+_s$ \\ 
 $[\mbox{GeV}/c]$ & $[\mbox{GeV}/c]$  &             &                       &                       &                   \\ \hline
 $5.5-6$      & $5.75$      & $ 7837\pm 220\pm 884$ &          ---             &      ---                 &            ---          \\
 $6-7$        & $6.5$       & $ 4056\pm  93\pm 441$ & $ 2421\pm 108\pm 424$ & $ 1961\pm  69\pm 332$ &               ---       \\
 $7-8$        & $7.5$       & $ 2052\pm  58\pm 227$ & $ 1147\pm  48\pm 145$ & $  986\pm  28\pm 156$ &                 ---      \\
 $8-10$       & $9.0$       & $  890\pm  25\pm 107$ & $  427\pm  16\pm  54$ & $  375\pm   9\pm  62$ & $ 236\pm  20\pm  67$ \\
 $10-12$      & $11.0$      & $  327\pm  15\pm  41$ & $  148\pm   8\pm  18$ & $  136\pm   4\pm  24$ & $  64\pm   9\pm  19$ \\
 $12-20$      & $16.0$      & $ 39.9\pm 2.3\pm 5.3$ & $ 23.8\pm 1.3\pm 3.2$ & $ 19.0\pm 0.6\pm 3.2$ & $ 9.0\pm 1.2\pm 2.7$ \\
\end{tabular}
\caption{Summary of the measured prompt charm meson
differential cross sections
and their uncertainties at the center of each $p_T$ bin.
The first error is statistical and the second systematic.
The products of the branching fractions~\cite{Hagiwara:fs} used are 
$(3.81\pm0.09)\%$, $(2.57\pm0.06)\%$, $(9.1\pm0.6)\%$ and $(1.8\pm0.5)\%$ 
for $D^0$, $D^{*+}$, $D^+$ and $D_s^+$, respectively.}
\label{table_Dx_D_bincorr}
\end{ruledtabular}  
\end{table}

\end{document}

%% file: Limited_Run2_Auth_revtex4.tex

\affiliation{Institute of Physics, Academia Sinica, Taipei, Taiwan 11529, Republic of China }
\affiliation{Argonne National Laboratory, Argonne, Illinois 60439 }
\affiliation{Istituto Nazionale di Fisica Nucleare, University of Bologna, I-40127 Bologna, Italy }
\affiliation{Brandeis University, Waltham, Massachusetts 02254 }
\affiliation{University of California at Davis, Davis, California 95616 }
\affiliation{University of California at Los Angeles, Los Angeles, California 90024 }
\affiliation{University of California at Santa Barbara, Santa Barbara, California 93106 }
\affiliation{Instituto de Fisica de Cantabria, CSIC-University of Cantabria, 39005 Santander, Spain }
\affiliation{Carnegie Mellon University, Pittsburgh, Pennsylvania 15213 }
\affiliation{Enrico Fermi Institute, University of Chicago, Chicago, Illinois 60637 }
\affiliation{Joint Institute for Nuclear Research, RU-141980 Dubna, Russia }
\affiliation{Duke University, Durham, North Carolina 27708 }
\affiliation{Fermi National Accelerator Laboratory, Batavia, Illinois 60510 }
\affiliation{University of Florida, Gainesville, Florida 32611 }
\affiliation{Laboratori Nazionali di Frascati, Istituto Nazionale di Fisica Nucleare, I-00044 Frascati, Italy }
\affiliation{University of Geneva, CH-1211 Geneva 4, Switzerland }
\affiliation{Glasgow University, Glasgow G12 8QQ, United Kingdom }
\affiliation{Harvard University, Cambridge, Massachusetts 02138 }
\affiliation{The Helsinki Group: Helsinki Institute of Physics; and Division of High Energy Physics, Department of Physical Sciences, University of Helsinki, FIN-00014 Helsinki, Finland }
\affiliation{Hiroshima University, Higashi-Hiroshima 724, Japan }
\affiliation{University of Illinois, Urbana, Illinois 61801 }
\affiliation{The Johns Hopkins University, Baltimore, Maryland 21218 }
\affiliation{Institut f\"ur Experimentelle Kernphysik, Universit\"at Karlsruhe, 76128 Karlsruhe, Germany }
\affiliation{High Energy Accelerator Research Organization (KEK), Tsukuba, Ibaraki 305, Japan }
\affiliation{Center for High Energy Physics: Kyungpook National University, Taegu 702-701; Seoul National University, Seoul 151-742; and SungKyunKwan University, Suwon 440-746; Korea }
\affiliation{Ernest Orlando Lawrence Berkeley National Laboratory, Berkeley, California 94720 }
\affiliation{University of Liverpool, Liverpool L69 7ZE, United Kingdom }
\affiliation{University College London, London WC1E 6BT, United Kingdom }
\affiliation{Massachusetts Institute of Technology, Cambridge, Massachusetts 02139 }
\affiliation{University of Michigan, Ann Arbor, Michigan 48109 }
\affiliation{Michigan State University, East Lansing, Michigan 48824 }
\affiliation{Institution for Theoretical and Experimental Physics, ITEP, Moscow 117259, Russia }
\affiliation{University of New Mexico, Albuquerque, New Mexico 87131 }
\affiliation{Northwestern University, Evanston, Illinois 60208 }
\affiliation{The Ohio State University, Columbus, Ohio 43210 }
\affiliation{Okayama University, Okayama 700-8530, Japan }
\affiliation{Osaka City University, Osaka 588, Japan }
\affiliation{University of Oxford, Oxford OX1 3RH, United Kingdom }
\affiliation{Universit\'a di Padova, Istituto Nazionale di Fisica Nucleare, Sezione di Padova-Trento, I-35131 Padova, Italy }
\affiliation{University of Pennsylvania, Philadelphia, Pennsylvania 19104 }
\affiliation{Istituto Nazionale di Fisica Nucleare, University and Scuola Normale Superiore of Pisa, I-56100 Pisa, Italy }
\affiliation{University of Pittsburgh, Pittsburgh, Pennsylvania 15260 }
\affiliation{Purdue University, West Lafayette, Indiana 47907 }
\affiliation{University of Rochester, Rochester, New York 14627 }
\affiliation{The Rockefeller University, New York, New York 10021 }
\affiliation{Instituto Nazionale de Fisica Nucleare, Sezione di Roma, University di Roma I, ``La Sapienza," I-00185 Roma, Italy }
\affiliation{Rutgers University, Piscataway, New Jersey 08855 }
\affiliation{Texas A\&M University, College Station, Texas 77843 }
\affiliation{Texas Tech University, Lubbock, Texas 79409 }
\affiliation{Institute of Particle Physics, University of Toronto, Toronto M5S 1A7, Canada }
\affiliation{Istituto Nazionale di Fisica Nucleare, Universities of Trieste and Udine, Italy }
\affiliation{University of Tsukuba, Tsukuba, Ibaraki 305, Japan }
\affiliation{Tufts University, Medford, Massachusetts 02155 }
\affiliation{Waseda University, Tokyo 169, Japan }
\affiliation{University of Wisconsin, Madison, Wisconsin 53706 }
\affiliation{Yale University, New Haven, Connecticut 06520 }


\author{D.~Acosta}
\affiliation{University of Florida, Gainesville, Florida 32611 }

\author{T.~Affolder}
\affiliation{University of California at Santa Barbara, Santa Barbara, California 93106 }

\author{M.H.~Ahn}
\affiliation{Center for High Energy Physics: Kyungpook National University, Taegu 702-701; Seoul National University, Seoul 151-742; and SungKyunKwan University, Suwon 440-746; Korea }

\author{T.~Akimoto}
\affiliation{University of Tsukuba, Tsukuba, Ibaraki 305, Japan }

\author{M.G.~Albrow}
\affiliation{Fermi National Accelerator Laboratory, Batavia, Illinois 60510 }

\author{D.~Ambrose}
\affiliation{University of Pennsylvania, Philadelphia, Pennsylvania 19104 }

\author{D.~Amidei}
\affiliation{University of Michigan, Ann Arbor, Michigan 48109 }

\author{A.~Anastassov}
\affiliation{Rutgers University, Piscataway, New Jersey 08855 }

\author{K.~Anikeev}
\affiliation{Massachusetts Institute of Technology, Cambridge, Massachusetts 02139 }

\author{A.~Annovi}
\affiliation{Istituto Nazionale di Fisica Nucleare, University and Scuola Normale Superiore of Pisa, I-56100 Pisa, Italy }

% ...
